\documentclass[conference,compsoc]{IEEEtran}
\ifCLASSOPTIONcompsoc
  \usepackage[nocompress]{cite}
\else
  \usepackage{cite}
\fi
\ifCLASSINFOpdf
\else
\fi

\usepackage{amsthm}
\usepackage{amsmath}
\usepackage{url}
\usepackage{amssymb}
\usepackage{fncychap}
\usepackage{algorithm}
\usepackage{mathtools}
\usepackage{subcaption}

\newtheorem{defi}{Definition}
\newtheorem{theorem}{Theorem}[section]

\newtheorem{remark}[theorem]{Remark}
\newtheorem{prop}[theorem]{Proposition}


\makeatletter
\def\BState{\State\hskip-\ALG@thistlm}
\makeatother

\begin{document}
%
\title{On the Relationship Between Modularity and Stability in Software Packages}

\author{\IEEEauthorblockN{Mohammad Raji}
\IEEEauthorblockA{Department of Electrical Engineering \& Computer Science\\University of Tennessee\\
EMail: mahmadza@vols.utk.edu}
\and
\IEEEauthorblockN{Behzad Montazeri}
\IEEEauthorblockA{Department of Computer Engineering \\ Razi University \\ EMail: behzad.montazeri@razi.ac.ir }}


%


\maketitle

\begin{abstract}
Modular and well-written software is an ideal that programmers strive to achieve. However, real-world project constraints limit the amount of reusable and modular code that programmers can produce. Many techniques exist that refactor code automatically using graph-based measurements and increase the quality and modularity of a codebase. While these measures work in the graph domain, their effect on the stability of software has been uncertain. In this work, we provide mathematical proof that modularity measures are indeed in favor of software stability. 
\end{abstract}


%
\IEEEpeerreviewmaketitle

\section{Introduction}
There are many properties that can be associated with good code. Sommerville describes good code as one that is highly maintainable, dependable, efficient and usable \cite{sommerville2004}. Truly reusable code is considered gold in the software industry as it significantly effects productivity and thus lowers costs \cite{lim1994} and without a doubt, good code is backed by a good design. However, in real-world scenarios, the great attributes of a good software might fade away as the project grows. Tight schedules, high customer demands and the high number of programmers involved in large projects are considered as some of the reasons that make efficient and engineered implementations change into a mess. 

Many refactoring and code analysis techniques have been introduced through the years. These methods are often based on modeling software components as a dependency graph \cite{wilde1991reusable}. Leveraging this model, many researchers have also introduced automatic software refactoring methods using various graph clustering and community detection algorithms \cite{dietrich2008cluster, pan2009class, alkhalid2011software, pan2013refactoring, raji2018refactoring}. 

While previous work have helped modularize code by increasing the cohesion inside software packages and decreasing the coupling between them with graph algorithms, their effect on software stability has not been formally analyzed. 

This paper studies the relationship between graph modularity and package stability and provides a mathematical proof that modularity is indeed in favor of more stable software packages. 

In the rest of this paper, Section \ref{sec:definitions} covers the definitions of software stability and graph modularity. Section \ref{sec:relationship} discusses the relationship between the two measures, and Section \ref{sec:conclusion} concludes the paper. 

\section{Definitions}
\label{sec:definitions}

\subsection{Software Stability}
\begin{figure}[ht!]
	\centering
	\begin{subfigure}[t]{0.49\columnwidth}
		\includegraphics[width=\columnwidth]{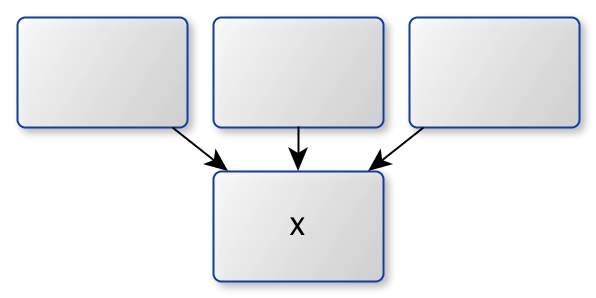}
		\caption{}
		\label{stable_package}
	\end{subfigure}
	\begin{subfigure}[t]{0.49\columnwidth}
	\includegraphics[width=\columnwidth]{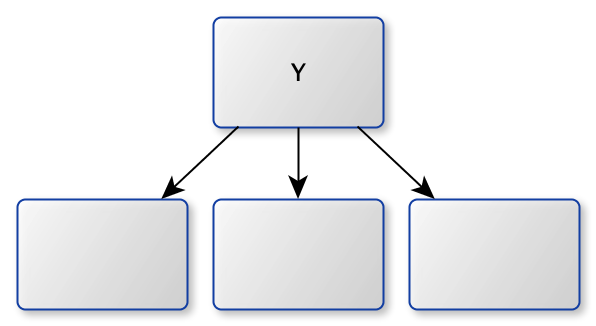}
		\caption{}
		\label{unstable_package}
	\end{subfigure}
	\caption{(a) shows a stable package that other packages depend on. (b) shows a very unstable package that depends on other packages only. }
\end{figure}

Robert Martin \cite{martin2003agile} defines stability as a measure proportional to ``responsibility''. A package is said to be responsible and independent if many other entities depend on it, while it doesn't depend on others itself. A package $p$ is said to be irresponsible and thus unstable, if it depends on many other entities, meaning that if they change, they cause $p$ to change as well. By Martin's definition, in Fig. \ref{stable_package}, X is an example of a stable package and in Fig. \ref{unstable_package}, Y resembles an unstable package. 

As a metric for stability, Martin defines the instability of a package as given in Eq. \ref{eq:Instability} where $I$ is instability, $C_a$ is afferent couplings and $C_e$ represents the number of efferent couplings. Afferent couplings is the number of classes outside the package that depend on classes within the package and efferent couplings is the number of classes within the package that depend on outside classes. 

\begin{equation}
\label{eq:Instability}
I=\frac{C_e}{C_a + C_e}
\end{equation}

If a package $p$ has an \textit{instability} of 0, then $p$ has maximum stability and if the package holds a value of 1 for \textit{instability}, then it would mean that the number of afferent couplings is 0 and therefore $p$ depends on other packages while no other package depends on $p$. This in turn would make it an extremely unstable package.

Martin also proposes the Stable Dependencies Principle (SDP) that helps the software design process by ensuring that modules that should be easily changeable not depend on modules that are harder to change \cite{martin2003agile}. In this case, packages should always have a higher $I$ metric than the ones they depend on. Consenting to this principle, one would be able to see a tree of packages, in which stable ones are placed at the bottom and the most unstable ones are at the top. The benefit of this approach is that packages that are violating SDP can be easily spotted. Any package depending on a package above it, would mean a violation of the principle. 

It is important to note that not all packages should or could be fully stable, as this would cause an unchangeable and inflexible system. Also, not all packages can be unstable as this would create an irresponsible system with a large number of connections and a high coupling. It is clear that pieces of code that are likely to change should be placed into unstable packages and pieces of code that are not very likely to change in the future should be placed in stable packages. Martin argues that high level design can not be placed in unstable packages because it resembles the architectural decisions of the projects, however if high level code is placed in stable packages then it would almost be impossible to change it after the project becomes more mature and more pieces of code start depending on it. The solution to this dillemma is the use of abstract classes that can introduce the flexibility and flow of stability that is needed. The basic idea behind the Stable Abstraction Principle (SAP) is that a package has to be as abstract as it is stable. This principle ensures that  the stability of a package does not contradict its flexibility. 

\subsection{Modularity}
Software dependencies have long been modeled as graphs. Software quality measures such as coupling and cohesion have direct meanings in the graph domain (inter-cluster, and intra-cluster relationships respectively). 
When using clustering or community detection for code refactoring, determining the quality of a clustering becomes extremely important. This measure should show how good a partition is. Quality functions associate a number with each cluster so that the clusters can be ranked and compared to one another. 

Arguably, the most common and famous quality function is Newman and Girvan's Modularity \cite{newman2004finding}, formally defining the meaning of modularity in graphs. Note that modularity has been a term that has been widely and loosely used in the software domain. Pan et al. use this graph definition of modularity in their automatic refactoring algorithm \cite{pan2013refactoring}. 

Modularity is based on the idea that a random graph contains no meaningful community. Based on this idea, if one can make a similar graph to the one being analyzed with the same number of vertices, edges and degrees but with edges placed at random, then by comparing it to the original graph one can find the major differences that have created communities. To understand the notion of modularity, we start by another measure for the goodness of a partition and build on it. Let $G$ be a graph with elements of its adjacency matrix presented as $A_{vw}$, where $A_{vw}$ is 1 if nodes $v$ and $w$ are connected and 0 otherwise, and $C_v$ being the community in which vertex $v$ belongs to. The following measure shows the fraction of edges in graph $G$, that fall within communities.

\begin{equation}
\label{eq:Fraction of edges in the same community}
\frac{\sum_{vw} A_{vw}\delta(C_v, C_w) } {\sum_{vw}(A_{vw}) } = \frac{1}{2m} \sum_{vw}A_{vw}\delta(C_v, C_w)
\end{equation}

where $\delta$ is the Kronecker delta function and $m$ is the number of edges in the graph. This fraction takes the value of 1 when all edges fall in one community and hence is not a good enough measure.

The idea behind modularity is that a random graph does not have a meaningful community structure and thus, if generated carefully, should provide a good point of comparison. Carefully generating a random graph that can depict the features and properties of the original graph but with no meaningful community is known as providing a null model in the area of complex systems. In this case, one can provide a graph which has the same amount of vertices, edges and vertex degrees while its edges are rewired randomly, so that the graph looses its community structure. In such a graph, the probability of an edge being in between vertices $v$ and $w$, if connections are made at random is calculated as below.

\begin{equation}
\label{eq:The probability of an edge between v and w}
\frac{k_vk_w}{2m}
\end{equation}

where $k_v$ and $k_w$ are the degrees of vertex $v$ and $w$ respectively. Now, by using equations \ref{eq:Fraction of edges in the same community} and \ref{eq:The probability of an edge between v and w}, one can calculate the modularity measure as

\begin{equation}
\label{eq:Modularity}
Q=\frac{1}{2m}\sum_{vw}[A_{vw} - \frac{k_vk_w}{2m}] \delta(C_v, C_w).
\end{equation}

By looking at Eq. \ref{eq:Modularity}, one can see some important aspects of this measure. The Kronecker delta function makes sure that a connection between two graph nodes in two different communities makes no contribution to modularity. Two connected nodes inside a community, make a positive contribution to modularity and the contribution is inversely proportional to the degrees of the two nodes. Also two nodes that are not connected, yet still reside in one community provide a negative contribution to the overall modularity of the clustering. 

\subsection{Directed Modularity}
For the stability measure to make sense, the dependency graph has to be directed. Therefore, we need to consider a directed version of the modularity measure as well. 

Several extensions of modularity for directed graphs have been proposed in the literature. Arenas et al \cite{arenas2007size} proposed an extension of modularity. Their idea is based on the fact that in a directed graph $G$, if vertex $i$ exists with more out-links and vertex $j$ exists with more in-links, then it is more probable that in a random rewiring a link be found from $i$ to $j$ rather than the opposite. Considering the original idea of modularity, this suggests that if an edge is found from $j$ to $i$, then this edge is contributing to a community structure more than $i$ to $j$ would, simply because it is more suprising and less random. By this definition, modularity can be altered for directed networks by changing the null model to a graph with the same number of vertices, edges, out-links and in-links as the original graph. The equation for modularity $Q$ in a graph with the adjacency matrix $A$ and $m$ number of edges can then be expressed as

\begin{equation}
\label{eq:Modularity_directed}
Q=\frac{1}{m}\sum_{ij}[A_{ij} - \frac{k_i^{out}k_j^{in}}{m}] \delta(C_i, C_j)
\end{equation}

where $\delta$ is the Kronecker delta function, $C_i$ and $C_j$ denote the communities that nodes $i$ and $j$ belong to, and $k_i^{out}$ and $k_j^{in}$ are the number of vertex $i$ and $j$'s out-links and in-links respectively.

\section{Relationship Between SDP and Modularity}
\label{sec:relationship}

In this section, the relationship between the directed version of modularity and the Stability Dependencies Principle (SDP) in refactoring packages is discussed. In a scenario where a package's class is chosen to be moved from one package to another using community detection methods, we show that modularity is in favor of SDP. We essentially show that hiding dependencies that violate SDP inside packages has a higher contribution to modularity than hiding non-violating dependencies. To show this behavior, some prior definitions are needed. 

\begin{defi}
	A movement of class $i$ from package $p_1$ to package $p_2$ is shown as the tuple $(i, p_1, p_2)$.
\end{defi}

\begin{defi}
	A border node in a package is defined as a node that has connections with nodes in other packages and thus directly effects the package's instability metric.
\end{defi}

SDP is generally satisfied in a case where no stable package depends on an unstable package. When considering the movement of only two border classes, while all other classes and packages are left intact, then the only dependencies effecting the two package's instability metric are the dependencies of the two border nodes. If a border node $i$ from stable package $p_1$ depends on a node $j$ from unstable package $p_2$, then clearly SDP is violated. 

\begin{remark}
	\label{sdp_satisfaction}
	Let $k_i^{out}$ and $k_j^{out}$ be the out-link degree of vertices $i$ and $j$ respectively, and $k_i^{in}$ and $k_j^{in}$ be the in-link degree of vertices $i$ and $j$. 
	If $k_i^{out} > k_i^{in}$ and $k_j^{out} < k_j^{in}$ and node $i$ and node $j$ are border nodes, then SDP is satisfied.
\end{remark}

\begin{remark}
	\label{sdp_dissatisfaction}
	Let $k_i^{out}$ and $k_j^{out}$ be the out-link degree of vertices $i$ and $j$ respectively, and $k_i^{in}$ and $k_j^{in}$ be the in-link degree of vertices $i$ and $j$. 
	If $k_i^{out} < k_i^{in}$ and $k_j^{out} > k_j^{in}$ and node $i$ and node $j$ are border nodes, then SDP is not satisfied.
\end{remark}

\begin{prop}
	\label{prop:sdp_mod}
	Let $i$ and $j$ be two classes in dependency graph $G$. If a movement $(i, c_i, c_j)$ exists and the conditions of remark \ref{sdp_satisfaction} holds, then the increase in modularity $Q$ is more, compared to the situation in which the conditions of remark \ref{sdp_dissatisfaction} holds true. 
\end{prop}

\begin{proof}
	Let $Q$ denote modularity while the conditions in remark \ref{sdp_satisfaction} holds true and $\bar{Q}$ denote modularity while the conditions in remark \ref{sdp_dissatisfaction} holds true. $Q$ and $\bar{Q}$ can be calculated using Eq. \ref{eq:Modularity_directed} as
	
	\begin{eqnarray*}
		Q&=&\frac{1}{m}\sum_{ij}[A_{ij} - \frac{k_i^{out}k_j^{in}}{m}] \delta(C_i, C_j), \\
		\bar{Q}&=&\frac{1}{m}\sum_{ij}[A_{ij} - \frac{\bar{k}_i^{out}\bar{k}_j^{in}}{m}] \delta(C_i, C_j).
	\end{eqnarray*}
	
	The bar on in-link or out-link $k$ denotes that it is being calculated in the scenario of remark \ref{sdp_dissatisfaction}, and is therefore equivelant to the out-link and in-link in the scenario of remark \ref{sdp_satisfaction} respectively. Thus one can write
	
	\begin{eqnarray*}
		\label{eq:kij_bar_relation}
		\bar{k}_i^{out} &=& k_j^{out} \\
		\bar{k}_j^{in} &=& k_i^{in}.
	\end{eqnarray*}
	
	By looking at the conditions of remark \ref{sdp_satisfaction} and remark \ref{sdp_dissatisfaction} it is clear that
	
	\begin{eqnarray*}
		\label{eq:kij_bar_relation2}
		\bar{k}_i^{out}\bar{k}_j^{in} &<& k_i^{out}k_j^{in} \\
		\frac{\bar{k}_i^{out}\bar{k}_j^{in}}{m} &<& \frac{k_i^{out}k_j^{in}}{m} \\
		A_{ij} - \frac{\bar{k}_i^{out}\bar{k}_j^{in}}{m} &>& A_{ij} - \frac{k_i^{out}k_j^{in}}{m} \\
		\bar{Q} &>& Q.
	\end{eqnarray*}
	
\end{proof}

The above proposition shows how modularity is compatible with the notion of SDP. Modularity is in favor of non-random structure in a network. Violating SDP would mean that a stable package is depending on an unstable package. In this scenario, the above proof shows that keeping two nodes that have violated SDP before, inside a single package is better for $Q$ than keeping two nodes that did not violate SDP. It is also important to note that if $i$ and $j$ belong to two different packages, then the condition will have no contribution to modularity and therefore is not discussed. 

\begin{figure}[ht!]
	\centering
	\includegraphics[width=.9\columnwidth]{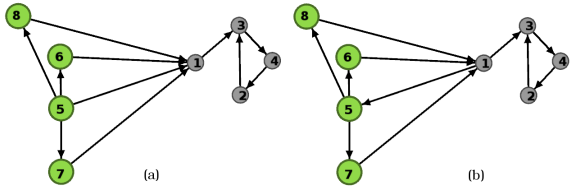}
	\caption{Two different graph dependency conditions.}
	\label{example1}
\end{figure}

As an example for the proved proposition, suppose that a system contains two packages $C_1$ and $C_2$, where $C_1$ is unstable and $C_2$ is a stable package. Two slighly different versions of this system is depicted in Fig. \ref{example1}. In both of these versions, vertices 1, 2, 3 and 4 are members of $C_2$ and vertices 5, 6, 7 and 8 belong to $C_1$. It is clear that in condition (b), edge $(1,5)$ is violating SDP. Based on Proposition \ref{prop:sdp_mod}, we show that moving node 1 from $C_2$ to $C_1$ has more positive contribution for package modularity, than in the case of condition (a). If movement $(1, C_2, C_1)$ happens, then four new edges positively contribute to the overall modularity of the dependency graph while one edge's contribution is eliminated. The reason for this is that edges between two communities provide no contribution to modularity because the kronecker delta function in Eq. \ref{eq:Modularity_directed} becomes zero. therefore edges $(5, 1)$, $(6, 1)$, $(7, 1)$ and $(8, 1)$ will have new contributions to modularity and edge $(1, 3)$ will no longer have any contribution. The changes in modularity $Q$ for condition (b) can be calculated using Eq. \ref{eq:Modularity_directed} as

\begin{eqnarray*}
	\label{eq:example1}
	\Delta Q &=& \overbrace{4(1 - \frac{1 \times 1}{2m})}^{\mathclap{\text{Contribution of the 4 new edges}}} - \underbrace{(1 - \frac{1 \times 1}{2m})}_{\mathclap{\text{Contribution of edge }(1, 3)}} = 3(1 - \frac{1 \times 1}{2m}).
\end{eqnarray*}

By replacing $m$ with the number of edges, we have

\begin{eqnarray*}
	\label{eq:example1b}
	\Delta Q &=& \frac{57}{20} = 2.85.
\end{eqnarray*}

Changes in modularity for condition (a) can be calculated the same way as follows.

\begin{eqnarray*}
	\label{eq:example1c}
	\Delta Q &=& \overbrace{(1 - \frac{4 \times 4}{2m})}^{\mathclap{\text{Contribution of edge }(5, 1)}} + 3(1-\frac{1 \times 4}{2m}) - (1 - \frac{1 \times 1}{2m}) \\& = & \frac{33}{20} = 1.65.
\end{eqnarray*}

The results clearly indicate that the graph gained more modularity when trying to suppress an SDP violation than when it is not. 

\section{Conclusion}
\label{sec:conclusion}
The use of dependency graphs in software refactoring has led to the usage of graph clustering and community detection methods for automatic placement of software components to increase their modularity. In this work, we looked at what modularity in software means for the stability of software packages. We provided a formal proof that modularity is in favor of the Stability Dependencies Principle. 



%
\bibliographystyle{IEEEtran}
\bibliography{IEEEabrv,main}

\end{document}